\documentclass[twoside]{article}

\usepackage[sc]{mathpazo} % Use the Palatino font
\usepackage[T1]{fontenc} % Use 8-bit encoding that has 256 glyphs
\usepackage{microtype} % Slightly tweak font spacing for aesthetics

\usepackage[hmarginratio=1:1,top=32mm,columnsep=20pt]{geometry} % Document margins
\usepackage{multicol} % Used for the two-column layout of the document
\usepackage[hang, small,labelfont=bf,up,textfont=it,up]{caption} % Custom captions under/above floats in tables or figures
\usepackage{booktabs} % Horizontal rules in tables
\usepackage{float} % Required for tables and figures in the multi-column environment - they need to be placed in specific locations with the [H] (e.g. \begin{table}[H])

\usepackage{lettrine} % The lettrine is the first enlarged letter at the beginning of the text
\usepackage{paralist} % Used for the compactitem environment which makes bullet points with less space between them

\usepackage{abstract} % Allows abstract customization
\usepackage{titlesec} % Allows customization of titles
\renewcommand\thesection{\Roman{section}} % Roman numerals for the sections
\renewcommand\thesubsection{\Alph{subsection}} % Roman numerals for subsections
\titleformat{\section}[block]{\Large\scshape\centering}{\thesection.}{1em}{} % Change the look of the section titles
\titleformat{\subsection}[block]{\large}{\thesubsection.}{1em}{} % Change the look of the section titles

\usepackage{fancyhdr} % Headers and footers
\usepackage[english]{babel}
\usepackage[utf8]{inputenc}
\usepackage{amsmath}
\usepackage{graphicx}
\usepackage[plainpages=false,pdfpagelabels]{hyperref}
\usepackage{cite}

\title{\vspace{-15mm}\fontsize{24pt}{10pt}\selectfont\textbf{Survey on Remote Electronic Voting}} % Article title

\author{
\large  
\textsc{Alexander Schneider} \qquad
\textsc{Christian Meter} \qquad
\textsc{Philipp Hagemeister}\\
\normalsize Heinrich Heine University D\"usseldorf \\ % Your institution
\normalsize \textit{firstname.lastname@uni-duesseldorf.de} % Your email address
\vspace{-5mm}
}
\date{}

\begin{document}

\maketitle % Insert title

\thispagestyle{fancy} % All pages have headers and footers

%----------------------------------------------------------------------------------------
%	ABSTRACT
%----------------------------------------------------------------------------------------

\begin{abstract}

\noindent Electronic and remote voting has become a large field of research and brought forth a multiplicity of
schemes, systems, cryptographic primitives as well as formal definitions and requirements for electronic elections.
In this survey we try to give a brief and precise overview and summary of the current situation.

\end{abstract}

%----------------------------------------------------------------------------------------
%	ARTICLE CONTENTS
%----------------------------------------------------------------------------------------

\begin{multicols}{2} % Two-column layout throughout the main article text

\section{Introduction}

Electronic voting sees a lot of adoption recently. Electronic voting systems are for example used in national-wide elections in Estonia \cite{madise2006voting}, 
for country council elections in Norway \cite{stenerud2012reality} and for canton wide votes in Switzerland \cite{gerlach2009three}.\\
To compete with traditional paper ballots, electronic voting system must ensure that the voter has the same or better privileges regarding
anonymity, integrity of the vote and many more. Electronic voting is not trivial and can not be compared with for example e-commerce, even if
it may come to mind. E-commerce has not found solutions for double spending and other fraud techniques and just recalculates
fees and prices accordingly - this is obviously not possible for e-voting. 
Because of the unique situation and requirements that electronic voting exists in it is quite hard to develop
an e-voting system that can be used in real-world elections and inspires trust in the voter. Our main contribution is the documentation of the
current state of e-voting research. 

In the following section properties which need to be fulfilled by modern and secure electronic voting systems are defined. In Section 3 possible attack vectors
are discussed. Sections 4 and 5 give an overview over existing e-voting schemes and systems. Open problems in e-voting are discussed in section 5 and section 6 gives
a brief overview of related work.

\section{Preliminaries}
Electronic voting systems claim to be at least as secure as traditional voting systems like paper ballots. In fact paper ballots (or even special voting machines) have many security issues. For example may the people counting the votes be corrupted and publish the wrong number of votes for a party. With the correct use of cryptography these issues can be eliminated, which is a great advantage for remote electronic voting systems. Therefore remote voting systems need to fulfill the following requirements:

\paragraph{Availability}
A voting system must remain available during the whole election and must serve voters connecting from untrusted clients.

\paragraph{Eligibility}
Only elective voters are allowed to cast one valid vote. Therefore no double votes are allowed.

\paragraph{Integrity}
The integrity of the vote must be guaranteed.

\paragraph{Anonymity and Election Secrecy}
The connection between the vote of a user and the user herself must not be able without her help.

\paragraph{Fairness}
Ensures that no (partial) results are published before the tallying has ended.

\paragraph{Receipt Freeness}
To reduce coercion, the user does not gain any information about her vote. Therefore she cannot proof her vote to anybody.

\paragraph{Correctness}
Election results must be counted properly and published correctly.

\paragraph{Robustness}
The system should be able to tolerate faulty votes.

\paragraph{Universal Verifiability}
After the tallying process, the results are published and can be verified by everybody.

\paragraph{Voter Verifiable}
The voter herself must be able to verify that her vote was counted properly.

\paragraph{Coercion}
Voting Systems must provide security aspects to prevent a coercer being able to force the voter to place a vote for a specific party, candidate etc.\\
In theory a voting system must be built coercion-resistant to guarantee that a voter can place her vote as intended even in the presence of a coercer. This implies that even sold or leaked credentials cannot be used to place ballots.

%%%%%%%%%%%%%%%%%%%%%%%%

\subsection{Usability}
Providing coercion-resistance and overall security leads on nearly all voting schemes / systems to a worse usability experience for the voter. Most of the cryptographic features need an action by the user, for example applying some audit codes to mark the correct vote or maintain a public / private key pair over the whole tallying process to authenticate at the system, place the vote and verify it at the end of the ballot.\\
Thus a voting system needs to be simple enough for the normal voter and ensure the requirements above.

\section{Attack Vectors}
There are a lot of different attack vectors which should be considered during the design of an e-voting system. We discuss some of the possible
attack vectors in this chapter.

\subsection{Infected Voter Machines}
One could easily think of malicious software on the voter machines which changes/deletes choices a voter makes during the voting procedure.
Widespread existing viruses, Trojan horses, etc. could easily be adapted to manipulate popular elections carried out via home computers or mobile devices.
Lifting the anonymity of the voter could also be one possible goal of the malicious software. 

\subsection{Active Network Attacks}
For a real world electronic election the standard network attacks like DDoS, Man-in-the-Middle, etc. have to be regarded. Since the election 
probably uses some kind of server-client architecture a single point of failure could be abused by an attacker. A Man-in-the-Middle
could also try to lift the anonymity of a voter if certain data is not encrypted.

\subsection{Network Correlation Attacks}

An attacker with ubiquitos Internet surveillance capabilities could forgo active attacks in favor of simply correlating the timing of communication and changes in the voting results, thereby lifting anonymity.

\subsection{Authorities Manipulating Votes}
There should be no possibility for any of the authorities to manipulate already cast votes without the voter or some auditor noticing it.
For example a corrupt authority could just not save an incoming vote or delete already saved votes. The authority could also try to change already cast votes
or use information about which voters did not use the opportunity to participate and vote in their stead.

\subsection{Double and Unauthorized Voting}
An attacker could try to vote without being authorized, which obviously should not be allowed. An authorized voter could also try to vote multiple times
in a way where all votes are counted. This should obviously be recognized and not allowed. 

\subsection{Coercion}
An adversary could use monetary or other motivation to persuade a voter into voting for a candidate the coercer wants. An e-voting system therefore has to
be mindful not to give any form of "voting receipt" to the voter. If the voter can not proof she voted like the coercer wanted  to even if the voter wants to prove it,
the coercer can not be certain that her manipulation will be successful and may henceforth abstain from coercion.
But this is not the only form of coercion. E-voting systems have to expect that voting credentials are sold or leaked to a coercer and used to vote in the voters stead.
This is very difficult to mitigate and an open problem regarding coercion.

\section{Schemes}

To date there are an astonishing number of different schemes describing the underlying cryptography of electronic voting systems.
Most of the schemes rely especially on some cryptographic primitive.
In the following section we will categorize some of them by the cryptographic primitives they mainly rely on.

\subsection{Group Signatures}
In a group signature every member of a group can sign a message on behalf of a group she is a member of.
Usually if a member signs twice her identity can be revealed through a process called \emph{tracing-by-escrowing}.
Another approach is \emph{tracing-by-linking} where double signatures can be traced without escrowing the identity of the double signer.

\cite{tsang2005short} propose a \emph{short linkable group signature} and its application to e-voting. Their scheme uses the group signatures during the tallying phase
where linked signatures are removed and every remaining signature is checked for validity before tallying the vote. This scheme however has the drawback to be coerceable.
The voter can simply give the coercer her authentication key and group signature, which the coercer uses to vote in the voters stead. Overall the proposed scheme is
only sketched out and the main contribution is the construction of the group signature.

Helbach et al. \cite{helbach2008code} extend an insecure code voting approach with linkable group signatures which uses \emph{tracing-by-escrowing}. Simply put, code voting is a procedure
where for every combination of voter and candidate a unique and random code is generated and used to vote for said candidate. 
Helbach et al. \cite{helbach2008code} argue that a code voting approach can be coerced by simply acquiring the voters code list. Their approach is to issue a
linkable group signature for every voter which must be used to sign the voting code. The authorities are
split into groups with different rights (e.g. one group which can issue the group signatures and one which can check if two signatures are linked by escrowing) 
to prohibit anonymity lifting.
We argue that this only prohibits the types of coercion where the voter is not aware of the coercion happening. If the voter wants to sell her vote or is bullied into
doing so, she can simply give away the code-list and her group-signature. 
Also if the signature-issuing and the group-checking authorities conspire anonymity could be possibly lifted. Helbach et al. further
stated that it is questionable to achieve universal verifiability with their system, because publishing the submitted voting codes could
give the coercer information about the success of her coercion attempts.

\subsection{Cryptographic Shuffles}
A cryptographic shuffle is intuitively understandable as the permutation of some elements in such a way that it is sufficiently random but still verifiable.
The verifiability is needed to proof the correctness of the shuffle. Typically a set of $n$ entities is responsible for the shuffling and the shuffle is secure as long
as at least one of the entities is not corrupted.

\cite{neff2001verifiable} construct efficient verifiable shuffles of $k$ DSA public keys and tuples. They propose a multi-authority scenario where the voter
encodes her vote through one or more ElGamal pairs and submits it to the authorities. The authorities then shuffle the votes and publish their proofs. Possibly other authorities
who have the power to decrypt the shuffled votes then decrypt and tally them. Since the shuffled votes are anonymized one could propose to publish the votes and the decryption key
to achieve universal verifiability. The voter has no means to check the integrity of her vote and coercion prevention is also not discussed.

\subsection{Homomorphic Encryption}
Homomorphic encryption is a type of encryption where operations on the encrypted text are reflected in the plaintext. For example: let $a = encrypt(2), b = encrypt(3)$ now 
$c= a \odot b$ could lead to $decrypt(c) = 5$. Which operations on the ciphertext translate to which on the plaintext is dependent upon the used encryption-algorithm. 
Exponential ElGamal and Paillier are two examples of popular homomorphic encryption algorithms. 

\cite{kiayias2004vector} propose the \emph{Vector Ballot e-Voting Approach}. Before the election $m$ authorities generate a public and private key for the election
according to a homomorphic encryption algorithm. The private key is jointly constructed by the $m$ authorities to achieve threshold encryption. In threshold encryption a 
threshold $t$ exists such that at least $t$ of the $m$ authorities are needed to decrypt a text encrypted with the corresponding public key.
The public key of the election is made public. Every voter constructs a vector consisting of entries corresponding to her choice. Each entry of the vector is encrypted with the
public key of the election. Beside the choice for predetermined candidates the vector contains a flag which indicates
whether a write-in candidate is chosen as well as the write-in candidate entry if one was chosen. The voter constructs a zero knowledge proof to show that her vector is well-formed.
Authorized voters then transmit their vote to the authorities. The authorities use heuristics to extract the write-ins while preserving anonymity. The non write-in votes
get compressed according to the homomorphic encryption that is used. Then at least $t$ authorities simply decrypt the compressed votes to determine the number of votes cast for the
predetermined candidates. The write-in portion gets send through a mix-net consisting of authorities and then opened separately for tallying.
All steps and the corresponding zero knowledge proofs are published on the bulletin board to guarantee universal verifiability. This scheme has no special coercion prevention mechanisms.

The coercion-resistant scheme by Spycher et al. \cite{spycher2012achieving} is based on the Coercion Resistant Scheme JCJ \cite{juels2005coercion}. 
Similar to the vector ballot scheme \cite{kiayias2004vector} Spycher et al. use ElGamal encryption, zero knowledge proofs and mix-nets. The authorities consist
of registrars and talliers. All communication is handled trough a bulletin board. 
The registrars construct real and fake credentials during the pre-voting phase. During the voting phase the voters can use her credentials
to vote normally. If a coercer demands from a voter to hand out her credentials, the voter can hand out fake credentials, which will be accepted by the bulletin board.
During the tallying phase the talliers remove all votes which have invalid proofs of correctness of fake credentials attached. Then the real votes are passed trough a mix-net,
decrypted and tallied. 

\subsection{Coercion Evidence}
There are many schemes which claim to achieve coercion-resistance or at least receipt-freeness, but to our knowledge all of them do so by requiring either a lot of
resources on the side of the voting officials or impend the usability for the typical voter by requiring her to perform tasks which are either too complicated or not realistic (e.g. 
providing untappable channels).
Grewal et al. \cite{grewal2013caveat} argue similarly and define \emph{coercion-evidence}. Coercion-evidence is 
fulfilled if the two conditions of \emph{coercion-evidence-test} and 
\emph{coercer independence} are met. No further requirements like an untappable channel or fake credentials are needed. The coercion-evidence-test condition is met if 
a tests exists which gets the public bulletin board as input and outputs an upper bound on the percentage of coerced votes. The coercer independence condition is met if
every honest voter follows the voting protocol normally.
Grewal et al. also propose a scheme they call \emph{Caveat Coercitor} that implements coercion-evidence and is based on the JCJ scheme \cite{juels2005coercion}.
In Caveat Coercitor multiple votes can be made and a heuristic, utilizing multiple votes of the same voter who chose different candidates, can determine coerced votes.
The benefit of that approach is that the coerced voter does not have to be aware of the coercion. Even coercion due to leaked credentials is successfully recognized
if the voter follows normal protocol for voting. The drawback is that enough voters who want to manipulate the election can choose not to follow the protocol and that way 
push the upper bound on coerced votes to a percentage where the election is rendered invalid. To solve this Grewal et al. propose to either disincentive dishonest voting by 
making it an offence or by replacing the coercion-evidence-test with a probabilistic variant, but leave the decision up to future work.

%%
%% TODO: subsubsections statt paragraphs bei B und C?
%%

\section{Systems}
Some complete voting systems exist today and were tested in several situations. Most of them use different cryptographic primitives as described below.

\subsection{Public Key Encryption}
Asymmetric key encryption uses two mathematically related keys (a \emph{key pair}) for encryption and decryption. One of those is kept private (the \emph{private key}) and the other one is made public (\emph{public key}). Those keys are typically created with the RSA algorithm, which can generate keys of size 2048 Kb for best usage and sufficient security. It is then possible to encrypt a message with the public key, decrypted with the private key and vice versa.\\
Most systems, using the Public Key Encryption, give the voter the choice on which computational device she wants to place the ballot. Because of the \emph{Strong RSA Assumption} \cite{boneh2011strong} the user can assume, the ballot is encrypted properly and there is no need for a special voting machine. Thus the voter can choose on which device she relies.

\subsubsection{DC Digital Vote-by-Mail Service}
The \emph{DC Digital Vote-by-Mail Service} (DVBM) \cite{stenbjorn2010vbm} is a voting system started in 2010 in Washington, D.C. as a pilot project by the \emph{Washington, D.C. Board of Elections and Ethics}. The main goal was to give the citizens, who are overseas during the election the possibility to vote over the web at the general election in November 2010.

\paragraph{Architecture}
DVBM is an open-source web application built with Ruby on Rails and is hosted on a typical Apache web server running MySQL as a relational database. The database stores encrypted ballots and all information about the elective users (including if a voter has already voted).

\paragraph{Ballot}
The voter authentificates with the system using her voter ID and a 16 character hexadecimal PIN number she received via postal mail. After the voter authenticated at the web application, she can download a PDF file where she can mark her vote. Then she must upload the file back to the web server, which encrypts it with the ballot's public key. The encrypted ballots are then stored in the MySQL database until the end of the election. On the last election-day a non-networked computer receives all ballots, decrypts them with the ballot's private key (which only this computer has) and tallies the votes.

\paragraph{Vulnerabilities}
During the public trial of DVBM, a team from University of Michigan found an exploit in a module used to upload the PDF files to the server \cite{wolchok2012attacking}. They gained access to nearly the complete server within 48 hours of the system going live. This case study was the first to analyze the security of a real election system.\\
Due to these vulnerabilities, the system was not used for the real election.

\paragraph{(Dis-)Advantages}
The system is built fairly simple and the applied cryptography is easy to understand. A drawback is that it is all built into one server rack and therefore a single point of failure exists as the attackers pointed out.\\
Furthermore concepts to prevent coercion do not exist and it needs an untappable channel for the authentication credentials.

%%%%%%%%%%%%%%%%%%%%%%%%%%%%%%

\subsubsection{Civitas}
\emph{Civitas} \cite{clarkson2008civitas} is a voting system, which provides coercion-resistance and is both universally and voter verifiable. It was developed by Clarkson, Chong and Myers from the Department of Computer Science, Cornell University. The system uses Public Key Encryption to encrypt the votes and some real and fake credentials for coercion-resistance. These credentials are generated the same way as it was in \emph{JCJ} and described above.

\paragraph{Architecture}
This architecture is split into several modules with distinct functions: The \emph{supervisor} sets up the election and publishes the ballot's public key. A \emph{registrar} publishes a list of all authorized voters with their identifier and public keys. Multiple \emph{registration tellers} authenticate the voters, give them their private credentials and generate a public key for a distributed encryption scheme. At the end multiple tabulation tellers collectively tally the election.\\

\paragraph{Ballot}
A voter must register at least partly in person, publish her public key to the registrar and must have a registration and a designation key. With these keys the voter can contact the registration tellers to run a specific protocol to get her private credentials. These can be used to place the encrypted vote and private credential to one of the distributed ballot boxes. The vote is replicated between the ballot boxes to prevent loosing the vote.\\
For tallying, the ballot boxes send all votes to the tabulation tellers, which decrypt the votes with the voters public keys and counts them. After evaluating a random number of votes, the tabulation tellers perform a random permutation, implemented with a \emph{mix-network} \cite{chaum1981untraceable} to guarantee anonymity.

\paragraph{(Dis-)Advantages}
Civitas achieves nearly all requirements for a feasible voting system. Though the implementation of coercion-resistance is fairly complicated, because the voter needs to create fake credentials, give them to the coercer, create real credentials and then vote with them. This lack of usability could implicate that coerced voters will not create new credentials and go through the voting process to have their votes counted.\\
The distributed multiple instances of some modules Civitas consists of, make it difficult to place fake votes or to drop votes. Even corrupting some of these instances is not enough to break this system, due to the distributed encryption scheme.

%%%%%%%%%%%%%%%%%%%%%%%%%%%%%%

\subsection{Homomorphic Encryption}

\subsubsection*{Web-based Open-Audit Voting}
\emph{Helios} \cite{adida2008helios} is an open-source web-based voting system built with modern web technologies. It does not provide any concepts for coercion-resistance and was built for low-coercion elections like ballots at a university.

\paragraph{Architecture}
Helios is implemented as a single web server and provides a single-page web application. In the back-end the administrator can manually add users and notify them via e-mail. In this mail the voter gets an automatically created 10-character password, which she can use to authenticate herself with the server.

\paragraph{Ballot}
This step differs from the other systems. The voter gets her credentials via mail including a link to the ballot and the ElGamal public ballot key. Following the URL leads directly to the options to vote for each sub-election. After all votes are made, the voter encrypts her choices with the public ElGamal key. Before submitting her vote, Helios adds some randomness to the encrypted vote and the voter has the possibility to review the choice to ensure all votes are set properly. If everything is fine, the voter can sign in with her username and password, seal the ballot and cast it to the server. If the audit option was chosen, the user can see the randomness added to the ballot and can see if the choices are correct. With the random factor and the choices, the voter herself can verify that everything was correctly encrypted. After this step, the ballot gets another randomness added to it and can be reviewed by the voter or is sealed and cast.\\
All votes are published on the server. For tallying the votes are shuffled and decrypted. Some proofs for correct shuffles are published so that everybody can verify the result of the ballot.

\paragraph{(Dis-)Advantages}
Helios is centralized and provides no mechanism against coercion, but is kept simple. Even reviewing the correct encryption of Helios is very easy to understand. Each step of the encryption is done in JavaScript, so the user can even disconnect the computer from the Internet after she downloaded all credentials for the ballot, do her choices, encrypt the vote and reconnect to the Internet to cast the ballot. Attacks, which need Internet connectivity, are therefore useless.

%%%%%%%%%%%%%%%%%%%%%%%%%%%%%%

\subsection{Blind Signatures + Secret Sharing}

\paragraph{Blind Signatures}
Assume Alice wants Bob to sign her message, but she does not want him to read this message. With blind signatures Alice would encrypt her message $m$ with her private key as known from RSA but add a random factor to this encryption and send the result $\tilde{m}$ to Bob. He may have Alice's public key, but because of the random factor he is not able to read this message. After Bob signs the message $sig(\tilde{m})$ and sends it back to Alice, she is able to remove the random factor and has a signature for her message $sig(m)$.\\
So Alice can get a valid signature for her message from Bob, but he was not able to decrypt her message.

\paragraph{Secret Sharing}
Secret sharing mechanisms described in \cite{shamir1979share, blakley1899safeguarding} can be used to divide a secret into pieces and distribute it to $n$ different entities. To reconstruct the secret, $t$ of $n$ entities need to cooperate.\\
In the context of electronic voting this threshold scheme can ensure that at least $t$ entities must be corrupted to gain access to the secret, e.g. the secret key of the ballot to decrypt the votes.

\subsubsection*{Secure Anonymous E-Voting System based on Discrete Logarithm Problem}
This web based voting system relies on blind signatures and shared secrets \cite{chen2014secure}. It provides anonymity through a net of proxy servers the voter needs to connect through to place her ballot.

\paragraph{Architecture}
The architecture is split into five modules: The \emph{Certificate Authority (CA)} where the user can get her personal certificate, an \emph{Authentication Center} to authenticate the user, some \emph{Public Proxy Servers} to hide the source IP of the voter, the \emph{Tally Center (TC)} which counts all votes and a \emph{Supervision Center (SC)} for supervising the tallying task.\\
TC and SC use \emph{secret sharing} to distribute their private keys for the tallying phase.

\paragraph{Ballot}
The voter needs to retrieve a personal certificate from the CA, which must be embedded into her web browser to authenticate at the Authentication Center and to retreive a blind signature from this entity over the voter's pseudonym the voter can choose. After the voter makes her voting decisions, her choices are encrypted with the public keys from the TC and the SC and send through the proxy servers to TC and SC.\\
After the tallying phase, the TC and SC cooperate to get their private keys to decrypt the ballots. Only with this cooperation it is possible to decrypt the ballots and count them correctly.

\paragraph{(Dis-)Advantages}
With this complex structure the system ensures integrity and anonymity, but provides no concepts for coercion. Secret sharing between TC and SC ensures integrity of the ballots, because $t$ out of $n$ TCs and SCs need to be fraudulent to corrupt the whole ballot. Most of the requirements are fulfilled, but the whole system needs many entities to work properly.

\section{Open Problems}
Despite the extensive research going on in the field of electronic and remote voting there are still a few open problems to consider.
In the following sections we name problems that are either not solved at all or just partially. The problems are not ordered by any criteria.

\subsection{Coercion-Resistance}
There are a lot of e-voting schemes and systems that tackle this particular problem \cite{grewal2013caveat,spycher2012achieving,bohli2007bingo,juels2005coercion}.
However there are problems with every one of the approaches. Some of the approaches have unrealistic requirements and are not suited for large-scale deployment usability-wise, because the
voter has to follow special protocols in case of coercion, most times of cryptographic nature. 
\cite{grewal2013caveat} mitigates this particular problem, but can only detect and not prevent coercion.
Overall there exists a lack of coercion-studies trying to analyze and measure the coercability in deployed real-world electronic and remote elections. One could think of
a study where one of the systems claiming to be coercion-free is deployed in a mock election and where one or more coercing elements try to successfully manipulate the election.

\subsection{Secure Platform Problem}
The term of the secure platform problem was introduced by \cite{rivest2001electronic}. The secure platform problem
describes the dilemma that the voter should be able to vote using a home computer or other device owned by the voter
but realistically can not, because it is easy to compromise said devices.
Most voting protocols assume some sort of trusted device to perform the voting on, but a lot of voter-owned devices
are either already infected with malicious software (up to 11 \% \cite{Malware}) or could easily be infected, if someone wants to manipulate an election.
Malicious software could change the voters choice, not send the vote, pretend to have voted correctly but misbehave, record the choice and lift the anonymity of the voter with the recorded ballot.
The list of possibilities is limitless. Furthermore the platform the election logic is running on (e.g. tallying-servers, bulletin boards, etc.) could be compromised as well.
There is almost no research devoted to design voting protocols that do not have a secure voting platform as a prerequisite.

\subsection{Usability and Comprehensibility}
The typical voter does not understand cryptography. She also uses her computer mostly for accessing the web and has no deep understanding
of computers and software. If this typical voter now gets confronted with electronic voting software that has a lot of options and is not designed with
usability as a primitive in mind the voter is most likely confused, overwhelmed and might even give up on trying to vote electronically.
Some systems already try to accomplish usability by running as a clearly structured step-by-step web application for example the Helios Voting System \cite{adida2008helios}.
But even if the application has good usability, a lot of voters do not understand and therefore do not trust the system. In the Norway electronic voting 
pilot \cite{stenerud2012reality} personalized return codes were used to send the voter a SMS confirmation of her vote. This was well understood by the voters
and heightened the trust in the system, although from a security standpoint the return codes did not contribute additional security.
From this we argue that usability and comprehensibility are quite important properties for a large scale voting system, that are often overlooked
by the researchers designing voting systems.

\subsection{Threat and Security Analysis}
Typically voting schemes and systems are analyzed regarding theoretical threats that could be possible because of weak cryptography or flaws in the protocol design.
But in real-world systems there is also the component of the implementation. A real-world election system is a piece of software running on servers connected to the Internet. 
As such they are potentially vulnerable to network-based attacks like DDoS, Man-in-the-Middle, packet sniffing, compromised key attacks, etc.
Another point of concern are the software-components used to implement the voting protocols. They could contain back-doors or use vulnerable libraries
thus potentially opening the whole system-network to attackers. 
As of now there is very little work analyzing those threats in regards to electronic voting. There are also no widely accepted and open frameworks that help implementing a secure electronic voting system or network.

\subsection{Distributed Systems}
Currently election systems are implemented using server-client architecture. As of default server-client architectures provide some kind of \textit{Single Point of Failure}.
There are systems like Civitas \cite{clarkson2008civitas} which divide the authorities into different parts (registrars, tellers, etc.), but still expose a
single point of failure. Imagine for example the registrars failing - suddenly nobody who is not already registered can register for the election.
We argue that completely distributed and still secure electronic voting systems are potentially possible and should be explored.

\subsection{Simple Distribution of Credentials}
This is a somewhat abstract point, because existing systems obviously are capable of distributing the voting credentials to the voters. But that typically happens over 
snail mail which is often insecure in itself. Further possibilities are credentials like the german ID card,
which contains an optional digital signature in the embedded RFID chip.

\section{Related Work}
Fouard et al. made an extensive comparison between specific schemes for electronic voting \cite{fouard2007survey} which currently seems to be rather unfinished. 
Lipmaa et al. \cite{lipmaa2005secure} is an article that gives an overview of different secure electronic voting protocols and 
J. Benaloh \cite{benaloh2013rethinking} published a study reviewing the state of coercion with focus on electronic voting and new technologies.\\
Our survey focuses on remote voting schemes and systems and gives a short overview about their cryptographic primitives.

\section{Conclusion}
In this survey we described current schemes and systems for remote electronic voting and gave an overview of possible attack-vectors and open problems.
We conclude that remote electronic voting is still in need of a lot of research and is not ready to be used in important large scale elections
like the presidential elections due to missing complete secure systems. In a low-coercion environment like student elections however remote electronic
voting is already an option.

\bibliographystyle{alphadin}
\footnotesize{
\bibliography{survey}}

\end{multicols}

\end{document}